# Finding Your Way Through the Jungle of Big Data Architectures

Torsten Priebe, Sebastian Neumaier
*Data Intelligence Research Group*
*St. Pölten University of Applied Sciences*, Austria
{torsten.priebe, sebastian.neumaier}@fhstp.ac.at

Stefan Markus
*Simplity AT GmbH*
Vienna, Austria
stefan.markus@simplity.ai

*Abstract*—This paper presents a systematic review of common analytical data architectures based on DAMA-DMBOK and ArchiMate. The paper is work in progress and provides a first view on Gartner's Logical Data Warehouse paradigm, Data Fabric and Dehghani's Data Mesh proposal as well as their interdependencies. It furthermore sketches the way forward how this work can be extended by covering more architecture paradigms (incl. classic Data Warehouse, Data Vault, Data Lake, Lambda and Kappa architectures) and introducing a template with among others "context", "problem" and "solution" descriptions, leading ultimately to a pattern system providing guidance for choosing the right architecture paradigm for the right situation.

*Index Terms*—Data Architecture, Data Lake, Logical Data Warehouse, Data Fabric, Data Mesh

## I. Introduction

Data science and machine learning are currently on everyone's lips. Since useful analyses are only possible on a solid data basis, architectures such as *Data Fabric* or *Data Mesh* are being proposed. But are those "modern" approaches really so different than "traditional" ones like *Data Warehouse*? The goal of this paper is a systematic review of the various (big) data architectures for analytics that have been proposed in the literature. In order to understand the similarities and differences, we propose a structure in ArchiMate notation [1] based on the key DAMA wheel elements [2].

Note that at this stage, the paper presents work in progress. We cover Gartner's *Logical Data Warehouse* and *Data Fabric* concepts [3, 4] as well as Dehghani's *Data Mesh* proposal [5]. Moving forward, we want to cover further architecture paradigms including classic *Data Warehouse* (both Kimball and Inmon style) [6, 7], Lindstedt's *Data Vault* [8]), *Data Lake* as well as *Lambda* and *Kappa* architectures [9, 10] and extend the model to clearly show the dependencies and shared elements as already indicated in figure 2 in the next section. This should ultimately lead to a pattern system similar to the GoF's software design patterns [11]. This pattern system will then provide guidance for choosing the right architecture paradigm for the right situation.

## II. Framework

As mentioned above, we base our structural framework on the DAMA-DMBOK and ArchiMate. We represent the DAMA elements *Data Integration & Interoperability*, *Data Storage & Operations*, *Data Quality*, *Data Security* and *Metadata* as ArchiMate application functions of a *Data Architecture*, which in turn is represented as an application collaboration. We decided to split the DAMA element *Data Warehousing & Business Intelligence* as we consider a data warehouse as a concrete application component in an according architecture. Furthermore, We amend *Business Intelligence* with *Data Science*, which is devoted a chapter in the DAMA-DMBOK, but did not make it into the wheel [2]. *Data Governance* and *Data Modeling & Design* are considered ArchiMate capabilities as they are not functions of the architecture, but rather supporting capabilities of the organization. We add *Data Sources* (as a data object) and *Data Consumers* (as a business actor) to the model and indicate the data flow as shown in figure 1.

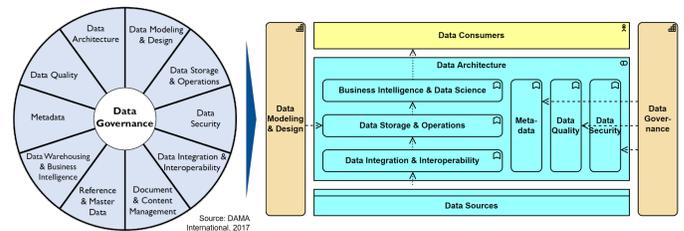

Fig. 1. Structural framework based on DAMA-DMBOK and ArchiMate.

Given the structure presented above, now consider the following data architecture paradigms: The (classic) *Data Warehouse Architecture* and the *Data Lake Architecture* both build the basis for Garnter's *Logical Data Warehouse Architecture*. The *Data Fabric Architecture* and its recent variation *Data Mesh Architecture* encorporate ideas from the *Lambda Architecture* and *Kappa Architecture* paradigms, which however focus mainly on the *Data Integration & Interoperability* function. The interdependencies are shown in figure 2. The following sections will cover the *Logical Data Warehouse*, *Data Fabric* and *Data Mesh* architectures in more detail.

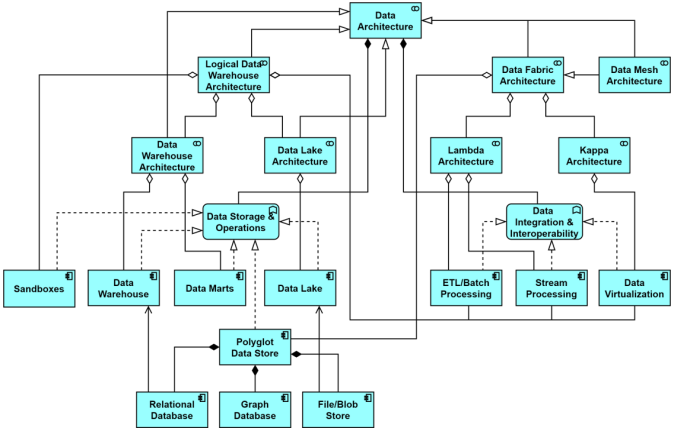

Fig. 2. Considered data architecture paradigms and their dependencies.

## III. Logical Data Warehouse

The *Logical Data Warehouse* concept was introduced by Gartner in 2012 [3] and provides recommendations how organizations can build a demand-driven, data management capability for analytical applications. According to the authors, architectural approaches such as *Data Warehouse*, *Data Lake* and *Data Virtualization* are not to be understood as concurrent solutions, but as complementary components of an overarching architecture. Gartner's Logical Data Warehouse Solution Path [12] also explicitly mentions *Sandboxes* and *Stream Processing* as components of a *Logical Data Warehouse Architecture*. This is shown accordingly in figure 3.

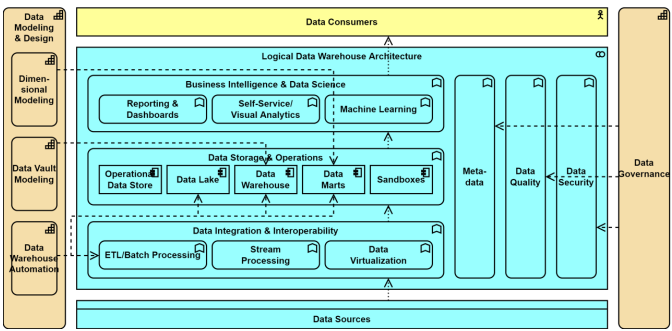

Fig. 3. Logical data warehouse architecture in our framework.

In terms of data models, Gartner mentions *Dimensional Modeling* and *Data Vault Modeling*. There is also a focus on *Data Warehouse Automation*, which is why we include those also as *Data Modeling & Design* capabilities.

## IV. Data Fabric

The combination of different data storage and integration techniques, however, without restriction to concrete architecture archetypes such as data lake or warehouse, led to the term *Data Fabric*, which was originally coined in 2015 by George Kurian of NetApp and then adopted and advocated again by Gartner [4]. According to the authors, *Data Fabric* is a design concept for attaining reusable and augmented data integration services, data pipelines and semantics for flexible and integrated data delivery.

*Data Fabric* may be seen as a successor and generalization of the *Logical Data Warehouse* concept presented in the previous section. It builds upon the idea of "polyglot persistence" [13] resp. a *Polyglot Data Store*, which combines storage approaches like *Relational Database*, *Graph Database* and/or *File/Blob Store* (such as Hadoop[1]). As mentioned before, it however does not promote concrete implementation paradigms like *Data Warehouse* or *Data Lake* as shown in figure 4.

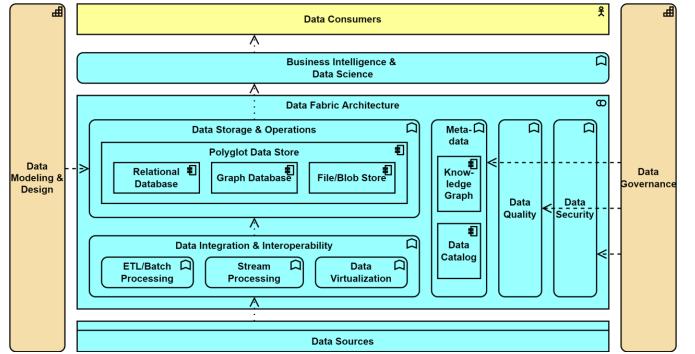

Fig. 4. Data fabric architecture in our framework.

The main focus area of the *Data Fabric* concept is *Metadata*, which according to Gartner consists of an (augmented) *Data Catalog* and a *Knowledge Graph* containing semantically linked metadata. The use of artificial intelligence resp. machine learning to partially automate the metadata creation are emphasized by Gartner's term "active metadata". On the contrary, reporting and analysis tools are not considered core scope of the *Data Fabric* concept [4], but rather under the responsibility of the *Data Consumers*. Therefore the *Data Fabric Architecture* box in figure 4 does not span the *Business Intelligence & Data Science* function.

## V. Data Mesh

The *Data Fabric* concept basically already introduces the idea of "data as a service" with datasets being "data products", even though this terminology is not explicitly used. This is what led in the end to the idea of a *Data Mesh* [5, 14]. The author argues that the existing centralized and monolithic data management platforms, with no clear domain boundaries and ownership of domain data, fail for large enterprises with a large and diverse number of data sources and consumers. In a *Data Mesh* the domains have to host and serve their datasets as domain data products, which enclose information and functionalities of the data. While the individual domain teams own the necessary technology to store, process and serve their data products, a common platform is needed to allow homogeneous interactions with the data products.

---
[1] https://hadoop.apache.org

Figure 5 shows this paradigm by adding *Business Domains* (represented as ArchiMate business collaborations) and their domain *Data Products* with their interfaces to the picture. As you can see in the diagram, *Data Pipelines* are also owned by the *Business Domains*, i.e. each domain is responsible for its own data transformations. A domain can consume data products from another domain. Like for *Data Fabric*, there is a strong focus on *Metadata* with a *Data Catalog* providing a cross-domain inventory of available data products. Also like for *Data Fabric*, reporting and analysis tools are not within the core scope of (and therefore *Business Intelligence & Data Science* is outside the *Data Mesh Architecture* box). However, unlike the other data architecture paradigms presented, the *Data Mesh* concept does take the *Data Sources* into closer consideration. Operational data is served via *Operational Data Products* (resp. their interfaces) just like *Analytical Data Products*.

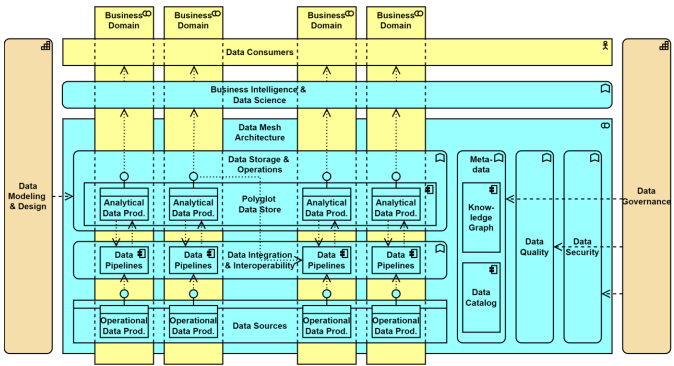

Fig. 5. Data mesh architecture in our framework.

## VI. Conclusion

In this paper we have provided a first systematic attempt of representing key (big) data architectures based on a common framework in a common semi-formal notation. We have used the DAMA-DMBOK and ArchiMate for this. In particular, this paper has covered *Logical Data Warehouse*, *Data Fabric* and *Data Mesh* architectures in more detail.

In future work, we want to cover further architecture paradigms such as *Lambda* and *Kappa* architectures, extend the model to clearly show the dependencies and shared elements – leading ultimately to a pattern system similar to the GoF's software design patterns [11]. This will involve introducing also other common elements of pattern templates like "context", "problem" and "solution", which will then provide guidance for choosing the right architecture paradigms.

## References

[1] The Open Group, *ArchiMate® 3.1 Specification*. van Haren Publishing, 2019.

[2] DAMA International, *DAMA-DMBOK: Data Management Body of Knowledge*, 2nd edition. Technics Publications, 2017.

[3] R. Edjlali and M. Beyer, "Understanding the Logical Data Warehouse: The Emerging Practice," Gartner, Tech. Rep. G00234996, 2012.

[4] E. Zaidi, E. Thoo, G. De Simoni, and M. Beyer, "Data Fabrics Add Augmented Intelligence to Modernize Your Data Integration," Gartner, Tech. Rep. G00450706, 2019.

[5] Z. Dehghani, *How to Move Beyond a Monolithic Data Lake to a Distributed Data Mesh*, 2019. [Online]. Available: https://martinfowler.com/articles/data-monolith-to-mesh.html (visited on 11/19/2021).

[6] R. Kimball and M. Ross, *The Data Warehouse Toolkit: The Definitive Guide to Dimensional Modeling*, 3rd edition. Wiley, 2013.

[7] W. H. Inmon, *Building the Data Warehouse*, 4th edition. Wiley, 2005.

[8] D. Linstedt and M. Olschimke, *Building a Scalable Data Warehouse with Data Vault 2.0*. Morgan Kaufmann, 2015.

[9] N. Marz and J. Warren, *Big Data: Principles and Best Practices of Scalable Realtime Data Systems*. Manning, 2015.

[10] J. Kreps, *I Heart Logs: Event Data, Stream Processing, and Data Integration*. O'Reilly Media, 2014.

[11] E. Gamma, R. Helm, R. E. Johnson, and J. Vlissides, *Design Patterns: Elements of Reusable Object-Oriented Software*. Prentice Hall, 1995.

[12] H. Cook, "Solution Path for Planning and Implementing the Logical Data Warehouse," Gartner, Tech. Rep. G00320563, 2017.

[13] M. Fowler, *Polyglot Persistence*, 2011. [Online]. Available: https://martinfowler.com/bliki/PolyglotPersistence.html (visited on 11/19/2021).

[14] Z. Dehghani, *Data Mesh Principles and Logical Architecture*, 2020. [Online]. Available: https://martinfowler.com/articles/data-mesh-principles.html (visited on 11/19/2021).